\documentstyle[aps,epsf]{revtex}
\begin{document}
\title{Rare isotope production in statistical multifragmentation}
\author{Scott Pratt, Wolfgang Bauer, Christopher Morling,  
and Patrick Underhill\footnote{Current address: Department
of Physics, Washington University, Campus Box 1105, One Brookings Drive, 
St. Louis, MO 63130}}
\address{Department of Physics and Astronomy and National Superconducting
Cyclotron Laboratory,\\
Michigan State University, East Lansing, MI 48824~~USA}
\date{\today}
\maketitle

\begin{abstract}
Producing rare isotopes through statistical multifragmentation is investigated
using the Mekjian method for exact solutions of the canonical ensemble. Both
the initial fragmentation and the the sequential decay are modeled in such a
way as to avoid Monte Carlo and thus provide yields for arbitrarily scarce
fragments. The importance of sequential decay, exact particle-number
conservation and the sensitivities to parameters such as density and
temperature are explored. Recent measurements of isotope ratios from the
fragmentation of different Sn isotopes are interpreted within this picture.
\end{abstract}

\pacs{25.70.Pq,24.10.Pa,64.60.My}

\section{Introduction}
The study of rare isotopes is attracting increasing attention due to the recent
development of radioactive beam facilities where isotopes are produced through
nuclear collisions. Projectile and target nuclei might vary in size from a few
dozen nucleons all the way to Uranium. The mechanism for rare isotope
production might entail the transfer of a few nucleons between the projectile
and target, induced fission of the projectile, or at highest energy transfer,
multifragmentation. This last mechanism, which assumes temperatures of a few
MeV, is the focus of this study.

Multifragmentation is an extremely complicated process that takes
place at time a time scale of 100 fm/c.  At such excitations tunneling
through barriers plays a less central role than in fission.  In this
short time the system is able to sample an enormous number of
configurations which makes dynamical calculations computationally very
intensive.  However, this very complexity lends justification to
statistical calculations which have proven remarkably successful in
predicting mass yields for excitation energies in the range of a few
MeV per nucleon \cite{smm,mmmc}.

It is our aim to apply simple statistical calculations to the study of the
production of rare isotopes, nuclei with neutron-to-proton ratios far from the
valley of stability. Our investigation addresses a variety of questions:

\begin{enumerate}
   \item How do yields depend on the physical parameters of the
   thermalized system?  Such parameters are size, overall neutron
   excess, density and temperature. 
   
   \item How sensitive are the yields with respect to various aspects
   of the modeling such as breakup density, level densities, and
   barrier penetration probabilities?
   
   \item Is there a qualitative and possibly quantitative explanation
   for the isospin amplification effects seen for light isotopes? 
   Furthermore, do these effects carry over into the production of
   heavier or exotic isotopes?  It has been experimentally observed
   that the ratio of mirror nuclei, e.g. $^{15}$O/$^{15}$N or
   t/$^3$He, can be of order 10 even though the neutron-to-proton
   ratio of the colliding nuclei is approximately 1.2
   \cite{snisotope_ratios}.
   
   \item By dividing the ratio of isotope yields from the
   fragmentation of $^{124}$Sn by the same ratio from the
   fragmentation of $^{112}$Sn, one can extract the relative chemical
   potentials between the two systems which appears to be robust with
   respect to sequential decay.  Can these results be quantitatively
   understood within the framework of a statistical model?
   
   \item Might multifragmentation provide a superior environment for the
   production of exotic isotopes in certain regions of the N/Z-plane?
   Currently, abrasion/ablation models \cite{cugnonschmidt,cugnon,lahet}
   provide the preferred scenario for creating rare isotopes and experiments
   have focused on searching for rare fragments at projectile rapidities.  The
   EPAX parameterization \cite{epax}, which is in common use for designing
   experiments, is built around such a picture.  It is not clear to what degree
   multifragmentative pictures can either complement or compete with
   abrasion/ablation models.
\end{enumerate}

The canonical ensemble is employed in this study.  The motivation for choosing
the canonical ensemble, along with descriptions of the statistical
fragmentation algorithm and the sequential decay calculation, appear in the
next section.  Isotope yields are primarily determined by the breakup
temperature and sequential decay.  These themes are introduced in Sec.
\ref{tempseqdecay_sec}.  The importance of enforcing exact conservation of
overall charge and baryon number with the canonical ensemble is discussed in
Sec.  \ref{grandvscanonical_sec}.  The sensitivity to the size and neutron
excess of the overall system is explored in Sec.  \ref{systemdependence_sec}
while Sec.  \ref{sensitivities_sec} contains a study of the sensitivities to
several aspects of the model such as breakup density, level density and
tunneling through the Coulomb barrier during the sequential decay process.
Recently, Xu et al.  have measured yields of light fragments from the
fragmentation of both $^{112}$Sn and $^{124}$Sn, and by taking ratios of
isotope yields have determined the relative chemical potentials of the two
systems \cite{snisotope_ratios}.  This result is interpreted with in
Sec.~\ref{deltamu_sec}.  A summary is provided in Sec.~\ref{summary_sec}.

\section{Modeling statistical multifragmentation and the subsequent decay}
\label{modeldescription_sec}

The initial statistical fragmentation of the system is calculated assuming the
canonical ensemble. The choice of the canonical ensemble is explained in the
following subsection. The recursive method described in
Sec.~\ref{canonicaldescription_subsec} provides a probability for creating any
state of any nuclide. The modeling of the subsequent sequential decay is
described in Sec.~\ref{seqdecay_subsec}.

\subsection{Choosing a statistical ensemble for fragmentation calculations}
\label{canonicalchoice_subsec}

Statistical calculations have proven remarkably successful in describing mass
yields for multifragmentative processes. Such models assume a fixed volume
where all distributions of $N$ neutrons and $Z$ protons into various fragments
and excited states are considered. A rather wide variety of thermal models have
been employed in the study of nuclear fragmentation, including grand canonical
\cite{fairandrup}, canonical \cite{smm} and microcanonical \cite{mmmc}
treatments. Additionally, isospin effects have been studied in lattice-gas
models \cite{pandasgupta,dasguptalatticegas,chomazlatticegas} and in
percolation \cite{kortemeyer,harreis}. Since our goal is to study the yields of
rarely produced isotopes which may be produced in future radioactive beam
facilities as rarely as in one per $10^{17}$ events, we employ the methods
recently promoted by Chase and
Mekjian\cite{chasemekjian,dasguptamekjian,dasguptapratt} which forego the need
of using Monte Carlo methods. Majumder and Das Gupta have in fact studied
Boron, Carbon and Nitrogen isotope production with a similar model to what is
presented here \cite{dasguptaisotopes}. The disadvantage of this method is that
explicit interaction of fragments (beyond mean field or excluded volume
effects) is outside the scope of the formalism.

Furthermore, we choose to focus on the canonical ensemble rather than the
microcanonical ensemble which would strictly constrain the energy of the
sampled configurations. The canonical distribution considers all configurations
of the same net $N$ and net $Z$, weighted by the Boltzmann factor, $e^{-\beta
E}$. The mean energy and variance of the energies are given by derivatives of
the canonical distribution,
\begin{eqnarray}
\langle H\rangle&=&-\frac{\partial \log(\Omega_c)}{\partial \beta}\\
\langle (H-\langle H\rangle)^2\rangle&=&
-\frac{\partial\langle H\rangle}{\partial\beta}\\
&=&T^2\frac{\partial \langle H\rangle}{\partial T}
\end{eqnarray}
It has been shown that these systems undergo first-order phase transitions,
which implies an infinite specific heat for infinite systems. However, for
systems of size $A\approx 200$ the energy rises by approximately 4 MeV per
nucleon as the temperature changes by one MeV in the range of $T\approx 6$ MeV.
This results in a variance of the energy per nucleon,
\begin{equation}
\sigma_{E/A}\approx \frac{2T}{\sqrt{A}}.
\end{equation}
Thus canonical calculations effectively sample configurations within an
excitation energy window of approximately 1.0 MeV per nucleon, or less if the 
systems are much larger than $A=100$, or if the temperature is away from the
transition region.

Unless an experiment gates on excitation energies to an accuracy better than
1.0 MeV per nucleon, there is little motivation to pursue a microcanonical
treatment. Microcanonical treatments have been shown to give much different
results than canonical treatments in the study of fluctuations and fragment
multiplicity distributions \cite{dasguptapratt}, topics which will not be
pursued in this study.

Grand canonical treatments sum over configurations with different total $Z$ and
$N$. Since we are interested in the production of extremely neutron-rich or
proton-rich fragments, it seems prudent to use the canonical ensemble. A
comparison of grand canonical and canonical calculations is presented in
section \ref {grandvscanonical_sec}.

\subsection{Fragmentation calculations with the canonical ensemble}
\label{canonicaldescription_subsec}

Our calculations are based upon the recursion relations of Chase and Mekjian
which permit one to sum over all partitions of the system into different
nuclear fragments. The recursion relation for the canonical distribution
function is
\begin{eqnarray}
\label{recursioneq}
\Omega_{Z,N}(T)&=&\sum_{c} e^{-\beta E_c}\\
\nonumber
&=&\sum_i \frac{a_i}{A}\,\omega_i(T)\,\Omega_{Z-z_i,N-n_i}(T),
\end{eqnarray}
where the label $c$ in the summation in the first line indicates one 
particular configuration, and $\omega_i$ is the partition function 
of a specific nuclear species $i$,
\begin{equation}
\label{omegadef_eq}
\omega_i(T)=\frac{V_{\rm red}(m_iT)^{3/2}}{(2\pi)^{3/2}}\sum_{{\rm
internal~levels}~j}e^{-\beta E_j},
\end{equation}
where $m_i$ is the mass and $z_i$, $n_i$ and $a_i$ are the charge neutron
number and baryon number of the species $i$. The reduced volume, $V_{\rm
red}=V_{\rm total}-2A/\rho_{b}$, roughly accounts for the overlap of the
nuclei. Given the default breakup density of one sixth $\rho_0$, the resulting
reduced volume is four times the volume of a system at normal nuclear density.

Calculating the partition functions of specific species $\omega_i$ is
straightforward given the levels and degeneracies of the nuclei.  This
was done for all fragments with $a<6$.  However, many of the nuclei
which are generated in this approach have not yet been observed, and
for almost all of them the level structure is not known.  Given our
lack of understanding of the ground state energies, let alone the
excited state energies, we employ a liquid-drop treatment.  We have
chosen the finite-range liquid-drop model (FRLDM) \cite{frldm} as a
means for generating ground state energies and have ignored the
microscopic terms in the model which account for shell structure.  The
spectrum of excited states was generated by assuming a uniformly
spaced assortment of single-particle states with spacing $\Delta
E=\alpha/\sqrt{A}$, where $\alpha$ is the level density parameter,
chosen to be 10 MeV. The degeneracy of a state with excitation energy
$n\Delta E$ was found by counting all ways to arrange particles and
holes such that they summed to the desired excitation energy.  For
heavy nuclei, the separation $\Delta E$ becomes small, and when the
separation fell below one MeV, the spectrum is interpolated onto a
mesh of 1.0 MeV resolution.  Finally, all mass-formula energies were
modified to account for screening of the Coulomb potential,
\begin{equation}
E_{\rm coul.}\rightarrow E_{\rm coul.}(1-\rho/\rho_0),
\end{equation}
where $\rho/\rho_0$ is the ratio of the density to nuclear matter density.

Once the functions $\omega_i(T)$ were calculated, the partition function
$\Omega_{N,Z}(T)$ was then generated through the recursion relation,
Eq. (\ref{recursioneq}), in less than a second of computer time. Subsequently,
the populations of individual levels $j$ of a species $i$ were found,
\begin{equation}
Y_{i,j}=\frac{d_{i,j}e^{-\beta E_{i,j}}\Omega_{Z-z_i,N-n_i}(T)}
{\Omega_{Z,N}(T)},
\end{equation}
where $E_{i,j}$ and $d_{i,j}$ are the energy and degeneracy of the level.

\subsection{Modeling sequential decay}
\label{seqdecay_subsec}

The vast majority of the nuclear levels considered in these calculations are
particle-unstable, including the ground states of those species outside the
proton and neutron drip lines. After the initial yields were calculated, the
subsequent decay was modeled by apportioning the weight of an unstable level
into all the levels into which the nucleus might decay. Eight decay modes were
considered: proton, neutron, deuteron, dineutron, diproton, t, $^3$He and
$\alpha$. The decay weights were chosen according to Weisskopf
arguments. Considering the decay from a state $i$ to a state $f$ via the
emission of particle of type $k$ with energy $E_k=E_f-E_i$,
\begin{equation}
\label{weisskopf_eq}
{\cal R}_{i\rightarrow f}(E_k)\propto mE_k\sigma_{k,f\rightarrow i},
\end{equation}
where $\sigma_{k,f\rightarrow i}$ is the capture cross section which we choose
as the geometric cross section, $\pi (R_f+R_k)^2$, scaled by the Coulomb
correction,
\begin{equation}
\label{sigmaclassical_eq}
\sigma_{k,f\rightarrow i}=\pi (R_f+R_k)^2\frac{E_k-E_{\rm coul}}{E_k}.
\end{equation}
Here, $E_{\rm coul}$ is the Coulomb barrier,
\begin{equation}
E_{\rm coul}=\frac{e^2Z_fZ_k}{R_f+R_k}.
\end{equation}

To account for particles tunneling through the Coulomb barrier, the
s-wave contribution to tunneling through the barrier was estimated through the
WKB approximation as
\begin{eqnarray}
\label{tunneling_eq}
\sigma_s(E)&=&\frac{\pi}{k^2}\exp\left\{-2\int_R^{r_0} 
      \sqrt{\frac{2m}{\hbar^2}\left(\frac{Z_fZ_ke^2}{r}-E\right)} 
      dr\right\},\nonumber\\ \nonumber 
&=&\frac{\pi}{k^2}\exp\left\{-\frac{2\sqrt{2\mu Er_0^2}}{\hbar}
\left[\arctan\sqrt{\frac{r_0-R}{R}}
-\frac{R}{r_0}\sqrt{\frac{r_0-R}{R}}\right]\right\},\\
r_0&=&\frac{Z_fZ_ke^2}{E}.
\end{eqnarray}
If the energy exceeded the Coulomb barrier, the capture cross section was taken
as $\sigma_s=\pi/k^2$. For all decays, both the classical form for
the cross section in Eq. (\ref{sigmaclassical_eq}) and the $s$ wave piece
described above were compared and the larger of the two cross sections was
chosen for the decay rates in Eq. (\ref{weisskopf_eq}).

Decays were calculated for all levels in all nuclei, beginning with the
heaviest nuclei. For the decay of each level, the decay rate was calculated
into every possible level energetically accessible through the eight decay
modes listed previously. The weight associated with the decaying nucleus was
then apportioned into all the states according to the rate for the decay
into each state. The weights were also simultaneously added into the ground
states of the eight nuclei representing the eight decay modes. Thus, the
decaying process exactly preserved the initial $N$ and $Z$ of the original
system. Given that heavier nuclei might have up to a thousand levels to decay
with each level decaying into any of several hundreds of levels in a daughter
nucleus, and given the existence of thousands of isotopes, the calculation of
sequential decays often required a few hours.

\section{Sensitivity to temperature and the role of sequential decay}
\label{tempseqdecay_sec}

Fragmentation is clearly sensitive to temperature, not only in the mass yields
but also in the isotopic yields.  Higher temperatures allow the system to
produce nuclei with unfavorable symmetry energies more efficiently.  However,
higher temperatures also result in hotter nuclei that experience longer decay
chains which in turn push the resultant residues closer to the valley of
stability.  The interplay of temperature and symmetry energy is discussed in
the following subsection by studying the liquid drop model.
Sec.~\ref{tempseqdecay_subsec} presents results of the canonical calculation
with and without sequential decay.  By modeling the fragmentation and decay of
a system with $A=200$ and $Z=80$, it is found that for lighter masses,
$a\approx 40$, isotopes with highest $n/a$ are best produced at temperatures
near 5 MeV. Heavier neutron-rich fragments, $a\approx 100$, are best produced
at lower temperatures near 3 MeV, but such production is certainly in the
domain of fission, and purely statistical calculations should not be taken too
seriously.

\subsection{Insights from the liquid-drop model}
\label{liquiddrop_subsec}

For a nuclear system at high excitation, where the initial nucleus is
nearly completely dissolved into the constituent protons and neutrons,
the neutron-to-proton ratio and the triton-to-$^3$He ratios are
roughly equal to the neutron-to-proton ratio of the decaying nucleus. 
At lower excitation, where most of the mass resides in large nuclear
fragments, this is no longer true.  In heavy nuclei, the relative
populations for nuclides of fixed $a$ are determined by the symmetry
energy, temperature and Coulomb energy.  Applying the grand canonical
ensemble, one can make a liquid-drop estimate of the relative
populations given the chemical potential,
$\mu_I\equiv(\mu_n-\mu_p)/2$, related to the difference in neutron and
proton number,
\begin{equation}
\label{liquiddrop_eq}
P_a(n-z)\propto\exp\left\{\left(-\xi(n-z)^2/a-b_c
\left[\frac{a+(z-n)}{2}\right]^2/a^{1/3}
+\mu_I(n-z)\right)/T\right\},
\end{equation}
where $\xi$ is the symmetry term in the liquid-drop model.  The
Coulomb term also contributes to the symmetry energy.  The parameters
chosen here are $b_c=0.7$ MeV and $\xi=23.4$ MeV. A more complicated
parameter set is employed in the FRLDM. However, for illustrative
purposes, we consider the Bethe-Weizs\"acker liquid-drop model here.

Ignoring the Coulomb term and completing the square in the exponential above
allows calculation of the mean neutron excess for fragments of mass $a$.
\begin{equation}
\langle n-z\rangle=a\frac{\mu_I}{2\xi}.
\end{equation}
The neutron to proton ratio is then,
\begin{equation}
\frac{Y_n}{Y_p}=\exp({2\mu_I/T})=\exp\left(4\frac{\xi}{T}
\frac{\langle n-z\rangle}{a}\right).
\end{equation}
If most nucleons reside in large fragments, the ratio $\langle n-z\rangle/a$
of a single species equals the ratio for the entire system.

By considering the limit where $\langle n-z\rangle/a$ is small, one can derive
the isospin amplification factor,
\begin{eqnarray}
\frac{Y_n-Y_p}{Y_n+Y_p}&=&\tanh \mu_I/T\\
&\approx&2\frac{\xi}{T}
\frac{\langle n-z\rangle}{a}.
\end{eqnarray}
Thus, the isospin amplification factor is $2\xi/T$, which exceeds 10 for
temperatures less than 5 MeV. This ratio is the same for all mirror nuclei
related by changing one proton to one neutron, e.g. $t-^3$He and
$^{15}$O-$^{15}$N. Large ratios of mirror nuclei have been observed in
intermediate-energy heavy ion collisions\cite{snisotope_ratios}. It should be
noted that Serot and M\"uller postulated similar ideas regarding isospin
fractionation in terms of phase separation \cite{serotmueller} and that
Samaddar and Das Gupta have made similar conclusions using the lattice gas
model \cite{dasguptalatticegas}.

The distribution described in Eq. (\ref{liquiddrop_eq}) can be described by an
offset and a Gaussian width. The offset is largely determined by the neutron
fraction $N/A$ of the composite system and does not depend strongly on the
temperature unless the temperature becomes so high that protons, neutrons,
tritons and $^3$He fragments can absorb much of any initial excess of the
neutron number.  The width depends on the ratio of symmetry energy to the
temperature.
\begin{equation}
\sigma_{n-z}=\sqrt{\frac{T a}{2\xi}}.
\end{equation}
Thus, although the ratio of yields of mirror nuclei tend to be largest for
lower temperatures, higher temperatures result in the broadest widths and the
greatest yields for nuclei with extreme neutron or proton imbalances. However,
this conclusion will be significantly modified by sequential decay.

\subsection{Results for the fragmentation of $A=200,~Z=80$ nuclei}
\label{tempseqdecay_subsec}

In this section we present results from the calculations described in
Section \ref{modeldescription_sec} which combine the canonical
ensemble with sequential decay.  We will see that the expectations
from the previous section, that higher temperatures produce more
isotopes far away from the valley of stability, are qualitatively
modified by the effects of sequential decay.  Sequential decay will
more strongly narrow the distribution when the temperature is higher
and more decays occur, allowing the final residue to approach the
valley of stability more closely.  Combined with the production
probability of producing a fragment of size $a$ with any $n-z$, the
optimum temperature for producing fragments with a large neutron
excess is not trivially understood.

\begin{figure}
\epsfxsize=0.6\textwidth 
\centerline{\epsfbox{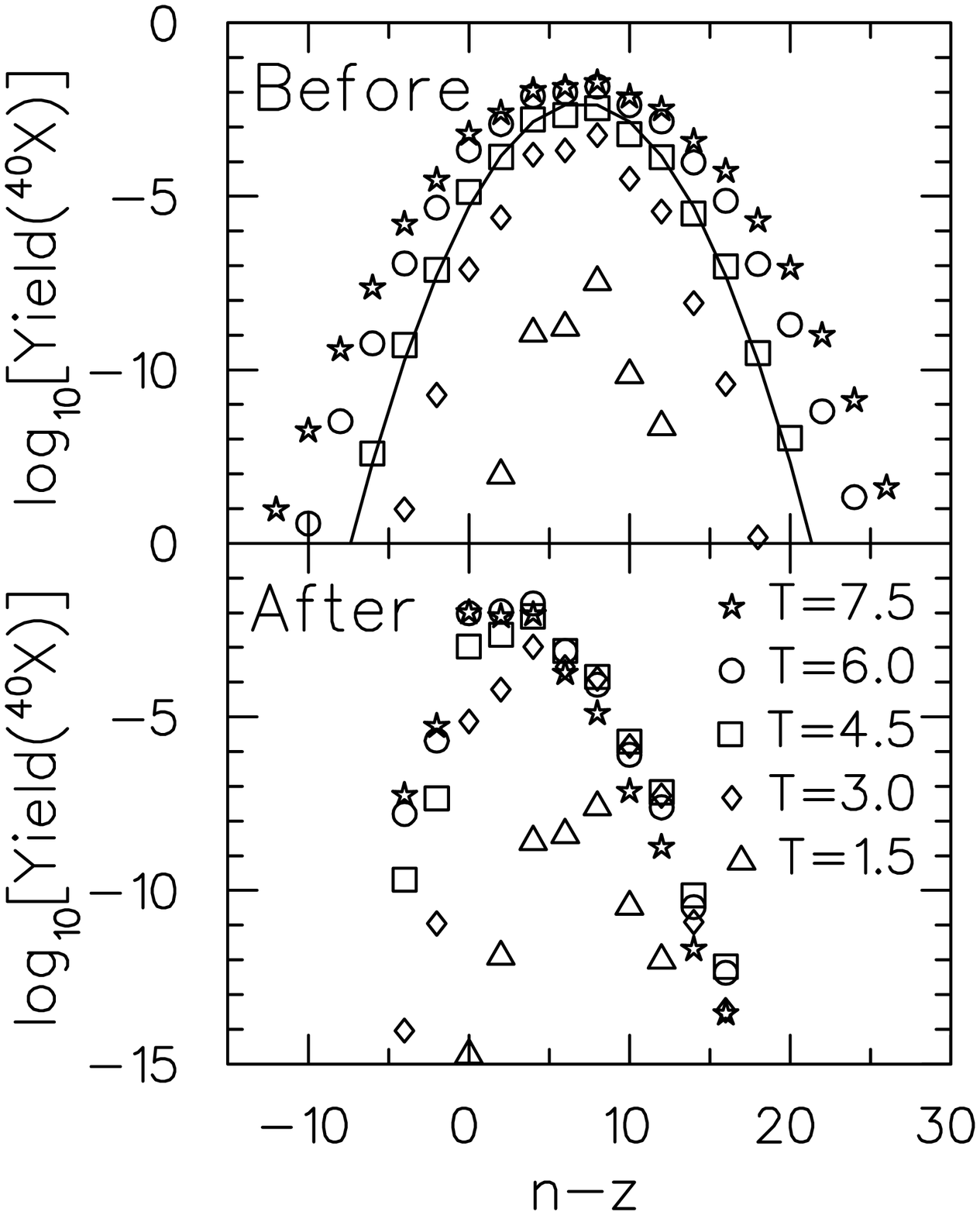}}
\caption{
\label{tempdependence_a200z80_a40_fig} 
Yields of mass 40 fragments are shown for the fragmentation of an $A=200$,
$Z=80$ system at a variety of temperatures. Yields are shown both for the case
where sequential decay is included (lower panel) and neglected (upper
panel). The pre-decay yields at 4.5 MeV are well described by a Gaussian (line)
where the width is determined by liquid drop parameters described in
Eq. (\ref{liquiddrop_eq}).  Pre-decay yield curves are broadest for higher
temperatures, but post-decay yields of very neutron rich fragments are largest
for temperatures near 5 MeV.}
\end{figure}

Figure \ref{tempdependence_a200z80_a40_fig} displays the yield of $a=40$
fragments as a function of $n-z$ for the fragmentation of an $A=200,~Z=80$
system at several temperatures. The upper panel displays yields for the case
where sequential decay is neglected. Clearly, higher temperatures create the
best conditions for creating highly neutron-rich nuclei when sequential decay
is neglected. The $T=4.5$ MeV results are well described by the Gaussian
resulting from the liquid-drop expression, Eq. (\ref{liquiddrop_eq}). The
magnitude and center of the Gaussian were adjusted to fit the peak but the
width was determined by the liquid-drop parameters. 

\begin{figure}
\epsfxsize=0.6\textwidth 
\centerline{\epsfbox{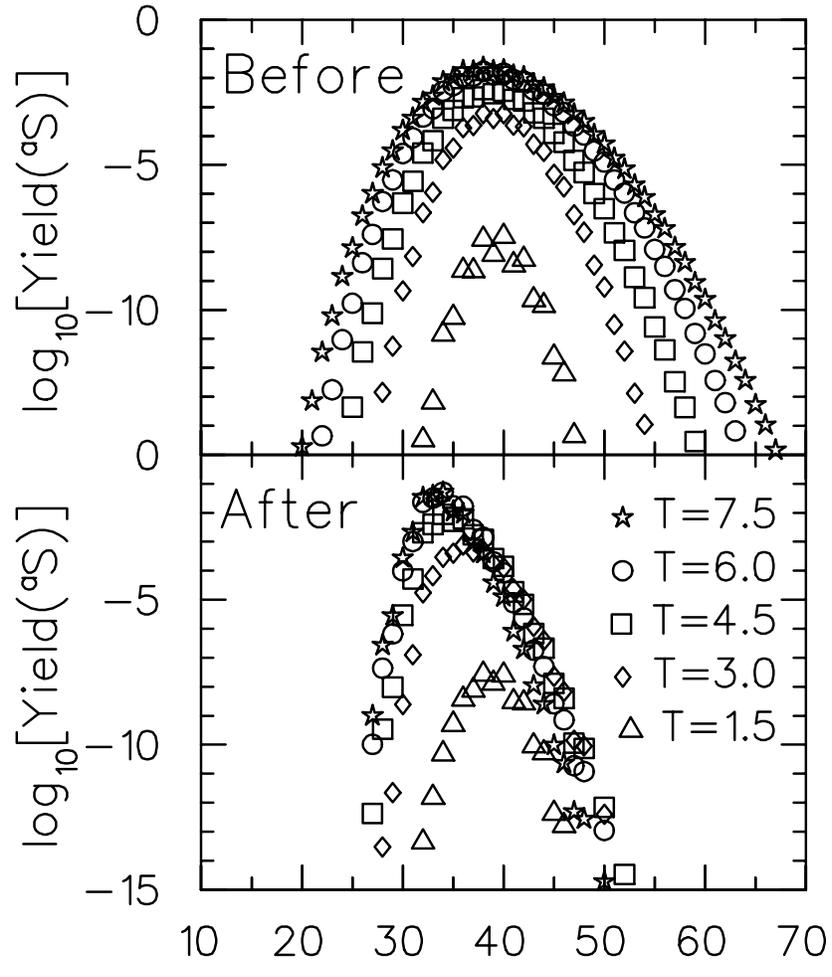}}
\caption{
\label{tempdependence_a200z80_sulphur_fig} 
Sulphur yields from the fragmentation of an $A=200$, $Z=80$ system are
displayed for a variety of temperatures. Yields are shown both for the case
where sequential decay is included (lower panel) and neglected (upper
panel). As in Fig.~\ref{tempdependence_a200z80_a40_fig} the best chance of
producing isotopes near the neutron drop line occurs for temperatures near 5
MeV.}
\end{figure}

Post-decay yields are displayed in the lower panel of
Fig.~\ref{tempdependence_a200z80_a40_fig}.  Sequential decay strongly
narrows the distribution, especially for higher temperatures.  Figure
\ref{tempdependence_a200z80_sulphur_fig} displays the yields of
sulphur isotopes for the same system.  From
Figs.~\ref{tempdependence_a200z80_a40_fig} and
\ref{tempdependence_a200z80_sulphur_fig}, one concludes that the
probability of producing rare neutron-rich isotopes is highest for
temperatures near 5 MeV for mass $a\approx 40$ fragments.

\begin{figure}
\epsfxsize=0.6\textwidth
\centerline{\epsfbox{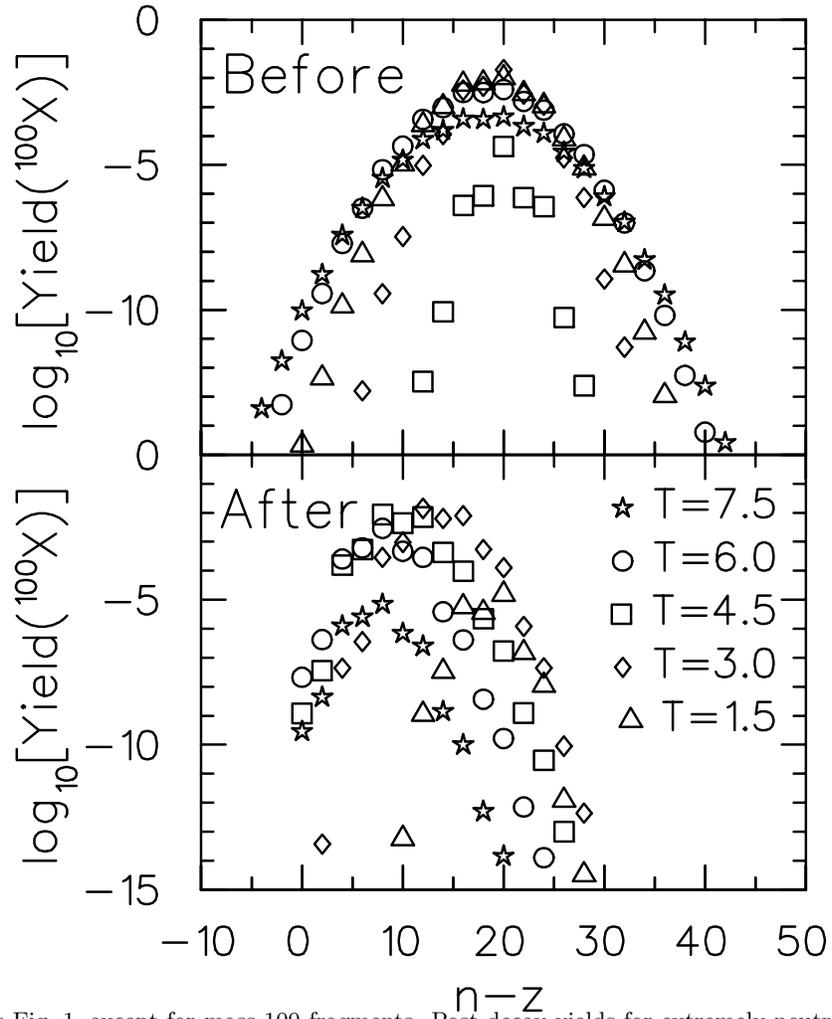}}
\caption{
\label{tempdependence_a200z80_a100_fig}
The same as Fig.~\ref{tempdependence_a200z80_a40_fig}, except for mass 100
fragments. Post-decay yields for extremely neutron-rich fragments in this mass
range are largest for temperatures near 3 MeV.}
\end{figure}

Yields of $a=100$ fragments are shown in
Fig.~\ref{tempdependence_a200z80_a100_fig}.  Since heavier fragments evaporate
more particles, they are more affected by sequential decay. For heavier
fragments, especially at temperatures less than 3 MeV, fission should provide
the dominant production mechanism. Although the statistical description
presented here does consider fission-like partitions of the system, both
dynamics and shell structure play a significant role in fragmentation at low
excitation where barriers play a pivotal role. For that reason, we confine our
future discussion to the production of lighter fragments, $a < A/3$.

\section{Grand canonical vs. canonical ensembles}
\label{grandvscanonical_sec}

The canonical ensembles employed in this study enforce exact conservation of
neutron and proton number. In order to understand the importance of this
constraint, we present corresponding calculations in the grand canonical
ensemble. In the grand canonical ensemble densities are determined by chemical
potentials, and the presence of a fragment of type $i$ does not affect the
possibility of observing a second fragment. A grand canonical ensemble supposes
either an infinite system or one chemically connected to an infinite bath of
both heat and particles.

Performing calculations in the grand canonical ensemble begins, as in the
canonical ensemble, by calculating partition functions as defined in
in Eq. (\ref{omegadef_eq}). The yield of fragments of type $i$ is then
\begin{equation}
Y_i=\omega_i(T)e^{\mu_nn_i/T+\mu_pz_i/T}.
\end{equation}
One must then find the chemical potentials $\mu_p$ and $\mu_n$ that give the
desired net neutron and proton numbers. In practice, this is accomplished
numerically by applying Newton's method. For the results presented here, only
nuclei of mass and charge less than that of the entire system were included in
the sum. Neglecting higher masses only affects results at low temperatures.

\begin{figure}
\epsfxsize=0.6\textwidth \centerline{\epsfbox{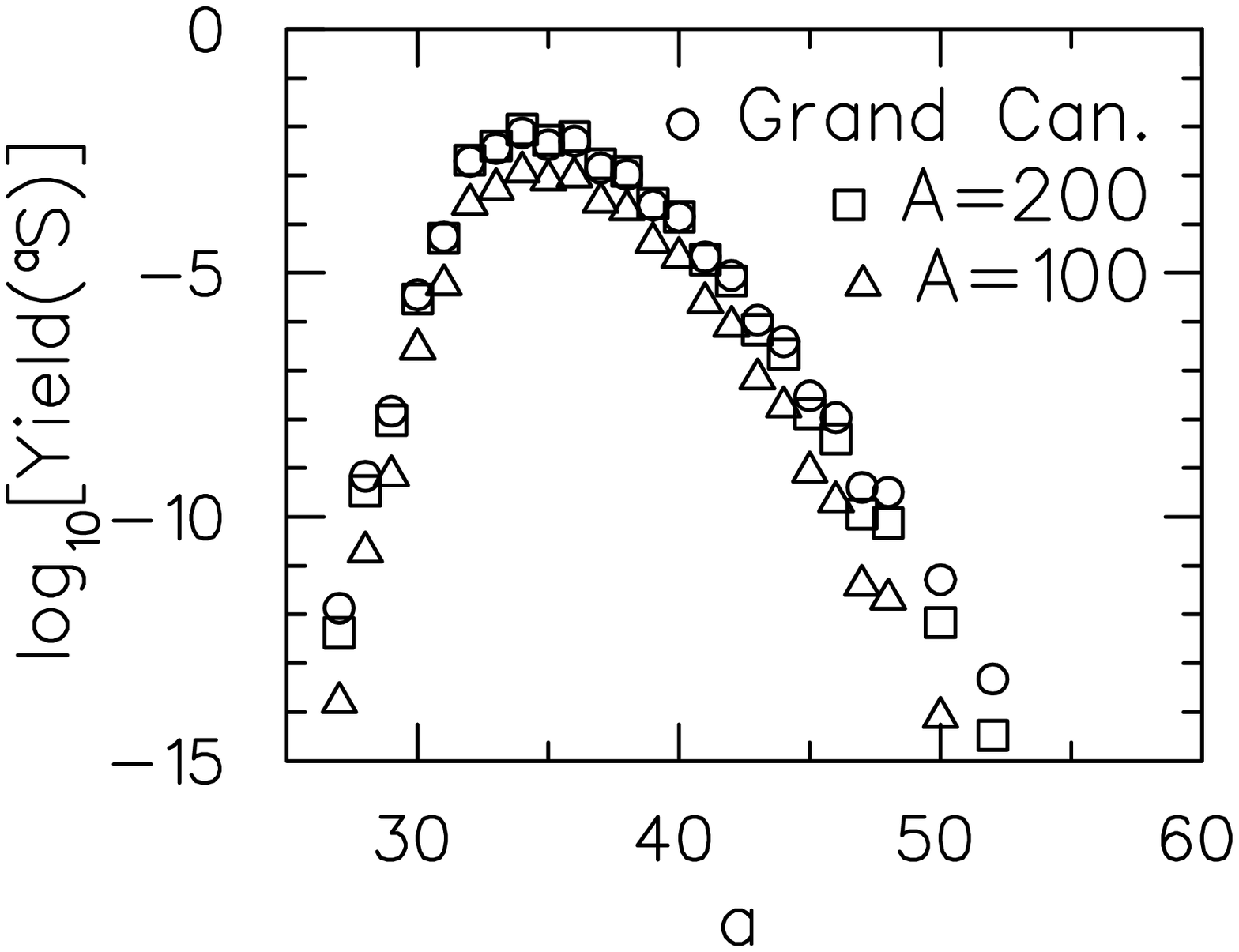}}
\caption{
\label{grandvscanonical_fig}
The importance of enforcing baryon number and charge conservation in the
statistical treatment is illustrated by comparing yields of sulphur isotopes
for canonical and grand canonical calculations.  As expected, fragmentation of
the $A=200,~Z=80$ system is better described by the grand canonical calculation
than fragmentation of the lighter system, $A=100,~Z=40$.}
\end{figure}

Yields of $a=40$ fragments from grand canonical and canonical calculations are
presented in Fig.~\ref{grandvscanonical_fig} for the fragmentation of an
$A=200,~Z=80$ system, and a system of half that size, and are compared to
results using the grand canonical ensemble. The yields are similar for the two
ensembles for the fragmentation of the heavier system, but are significantly
different for the lighter example. The reduced production of extremely
neutron-rich nuclei from the fragmentation of the $A=100$ system is easily
understood as resulting from the absolute conservation of neutron number. In
summary, use of the grand canonical ensemble is invalid for smaller systems,
and even in the case of large systems, can result in overpredicting yields by
an order of magnitude near the neutron drip lines.

\section{Isotope production as a function of system size and charge}
\label{systemdependence_sec}

By fragmenting heavier systems, one is able to create hot systems with
higher neutron fractions.  By choosing larger systems, yields are less
affected by the constraints of exactly conserving the net number of
neutrons and protons.  Although the qualitative trend is clear --
larger and more neutron-rich systems are better suited to producing
extremely neutron-rich fragments -- quantitative dependencies might be
rather surprising given the amplifications mentioned earlier.

\begin{figure}
\epsfxsize=0.6\textwidth \centerline{\epsfbox{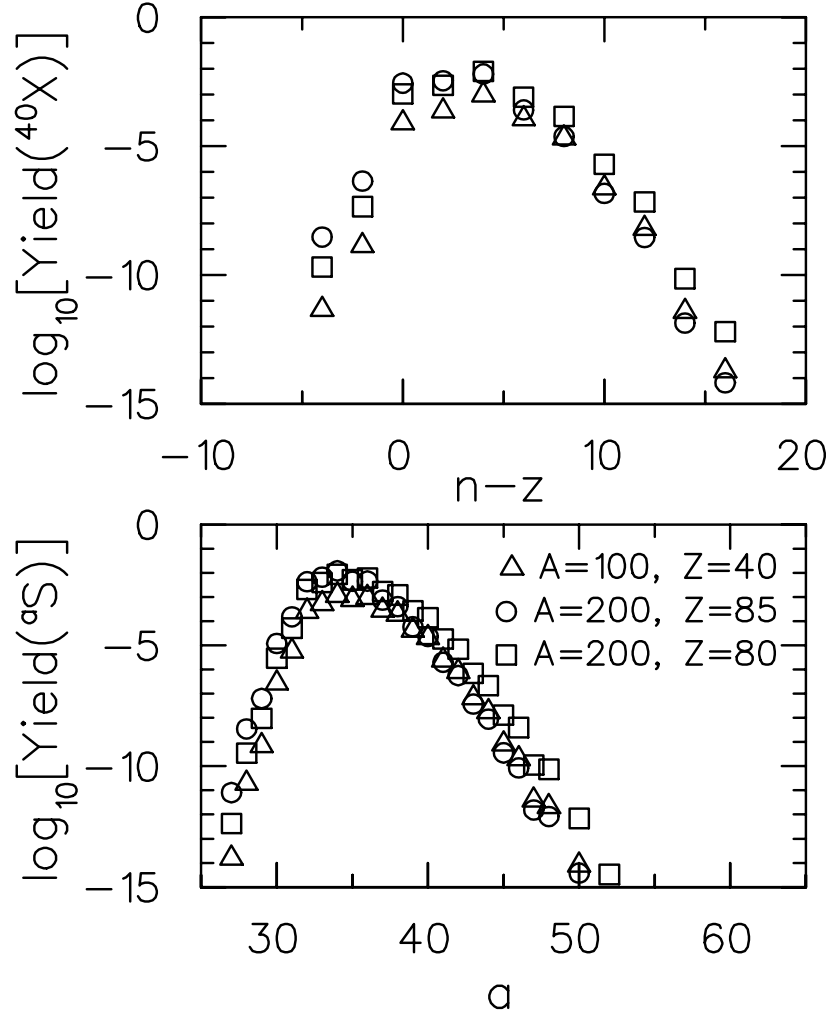}}
\caption{
\label{systemdependence_fig}
Yields of mass 40 fragments (upper panel) and sulphur isotopes (lower panel)
are shown for a variety of systems fragmenting at a temperature of 4.5
MeV. Yields of neutron-rich isotopes were moderately decreased when considering
a smaller overall system of the same neutron fraction. For the $A=200$ systems
yields were substantially reduced by lowering the net number of neutrons from
120 to 115.}
\end{figure}

Yields for sulphur fragments are shown in
Fig.~\ref{systemdependence_fig} for three systems: $(A=200,Z=80)$,
$(A=200,Z=85)$, and $(A=100,Z=40)$.  Each system is chosen to fragment
at a temperature of 4.5 MeV. The importance of choosing the most
neutron-rich isotope is clearly apparent in the comparison of the
$Z=80$ and the $Z=85$ systems.  Here, the neutron fraction changes
from 0.6 to 0.575.  Furthermore, comparison with the smaller system at
the same neutron fraction $N/A=0.6$ demonstrates the modest
enhancement of neutron-rich fragments involved in using larger
systems.

Given the sensitivity of the yields to changing the neutron and proton number
of the fragmenting system, evaporation of neutrons during the early stages of
the reaction should significantly reduce the production of rare neutron-rich
isotopes. During the first 100 fm/c, Weisskopf arguments suggest that 20 to 30
neutrons, but only 5 to 10 protons, might evaporate from a Uranium nucleus
heated to a temperature of 10 MeV.  The neutron fraction of the residue might
thus change from 0.614 to $\approx$0.59 during the evaporation which
significantly reduces the possibility to emit fragments near the neutron drip
line. Fission yields are also affected by the number of evaporated pre-fission
neutrons, which can be sizable, especially when the fission time scale is
large.

If one wishes to find reactions with the highest abundances of rare
neutron-rich fragments, one should try to minimize the number of neutrons
emitted in the pre-fragmentation stage of the collision.  Reducing the number
of neutrons emitted during the pre-fragmentation stage might be
accomplished by choosing heavy-ion collisions as a means of producing
multifragmented systems. Since heavy ion collisions, especially mid-central
collisions, retain some of the initial collective longitudinal motion of the
beam, the expansion does not require as much time to accelerate from rest as it
would in a $pA$ collision. Furthermore, by colliding two very heavy nuclei, the
global system would be relatively proton rich and not emit neutrons as
preferentially, relative to protons, during the early stages of the
collision. In fact, the mid-rapidity region, often referred to as the neck,
might be even more neutron rich than the target or projectile
\cite{indraplagnol}. Of course, the disadvantage to using heavy-ion collisions
is that lower beam energies are used and the produced fragments are not as
forward focused which might make them more difficult to detect. If one wishes
to produce fragments with charges greater than Calcium, $z=20$, and energies 
less than 30$a$ MeV, electron recombination and the resulting impure charge
states makes detection and refocusing especially difficult.

\section{Sensitivity to breakup density, level density and sub-barrier 
evaporation}
\label{sensitivities_sec}

Although the calculated yields of rare fragments depend on details of the
modeling procedure, it is not obvious to what degree yields are sensitive to
particular aspects of the modeling.  One obvious parameter is the symmetry
energy parameter.  As the symmetry energy is uncertain by one or two MeV
outside the region of measured nuclei the associated Boltzmann factors might
change by a factor of two.  Clearly, a lower symmetry energy will lead to
broader isotope distributions.  The sensitivity to other aspects of the
modeling is not so trivial.  In this section we consider three sources of
uncertainty: the breakup density, the level density parameter and sub-barrier
evaporation.

\begin{figure}
\epsfxsize=0.6\textwidth 
\centerline{\epsfbox{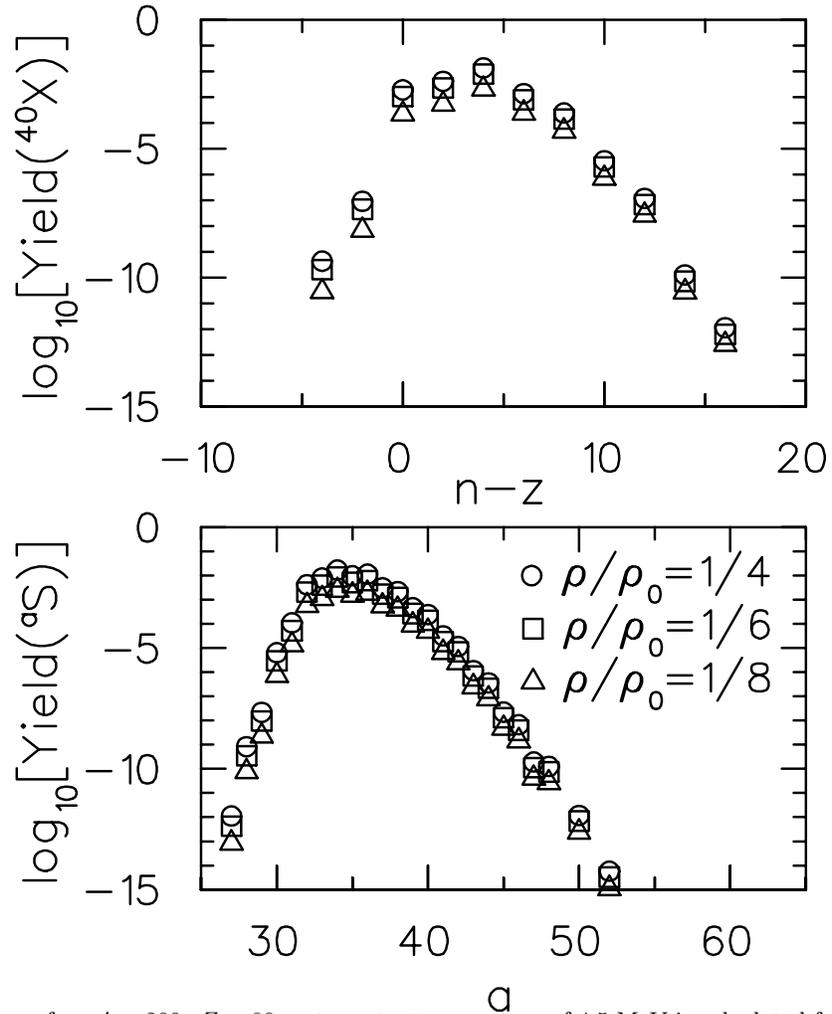}}
\caption{
\label{densitydependence_fig}
Fragmentation of an $A=200,~Z=80$ system at a temperature of 4.5 MeV is
calculated for three different breakup densities. Yields are shown for mass 40
fragments as a function of $n-z$ (upper panel) and for sulphur fragments as a
function of mass (lower panel). The breakup density affects the mass
distribution more strongly than the relative probabilities for isotopes of
fixed mass.}
\end{figure}

Adjusting the breakup density affects the mass distributions as larger
breakup volumes lead to preferential emission of many small fragments. 
To determine the sensitivity of the isospin distributions to the
breakup density, the fragmentation of an $A=200, Z=80$ system was
considered at three breakup densities, $\rho_0/8$, the default breakup
density $\rho_0/6$, and $\rho_0/4$.  The yield of fragments of mass
$a=40$ and of sulphur isotopes are displayed in
Fig.~\ref{densitydependence_fig} assuming a breakup temperature of 4.5
MeV. The distributions of isotopes of fixed mass are fairly
insensitive to the choice of breakup density.  This is not surprising
given the liquid-drop arguments presented in
Sec.~\ref{liquiddrop_subsec}.  The distribution of sulphur isotopes is
broader for the higher breakup densities as would be expected since
higher densities lead to increased production of heavier fragments.

\begin{figure}
\epsfxsize=0.6\textwidth
\centerline{\epsfbox{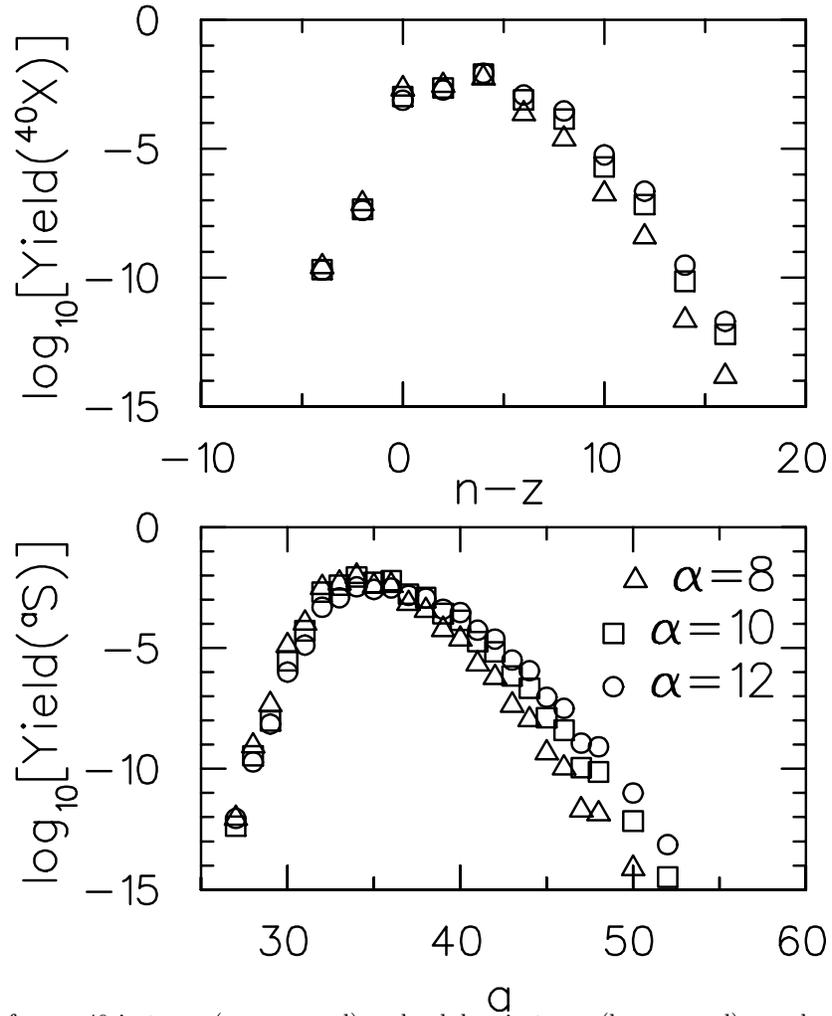}}
\caption{
\label{leveldensdependence_fig}
The yields of mass 40 isotopes (upper panel) and sulphur isotopes (lower panel)
are shown to be sensitive to the choice of level density parameter
$\alpha$. Higher level density parameters lead to broader isotope distributions
due to the dynamics of sequential decay. Again, calculations concerned an
$A=200,~Z=80$ system at a temperature of 4.5 MeV.}
\end{figure}

The level density parameter affects results with regards to the
initial population of isotopes, the relative populations of levels of
a given isotope, and in the dynamics of the sequential decay.  The
breakup of an $A=200,~Z=80$ system is again considered, and the yields
of mass 40 fragments and sulphur isotopes are displayed in
Fig.~\ref{leveldensdependence_fig} for three level density parameters,
8 MeV, the default value of 10 MeV, and 12 MeV. The distributions are
noticeably broader for the higher level density parameters.  Very
little effect was evident in the pre-decay yields and thus this
sensitivity derives from the dynamics of the sequential decay.  This
suggests the importance of improving the modeling the density of
excited states as the level density is known to have a strong
dependence on shell structure.

\begin{figure}
\epsfxsize=0.6\textwidth
\centerline{\epsfbox{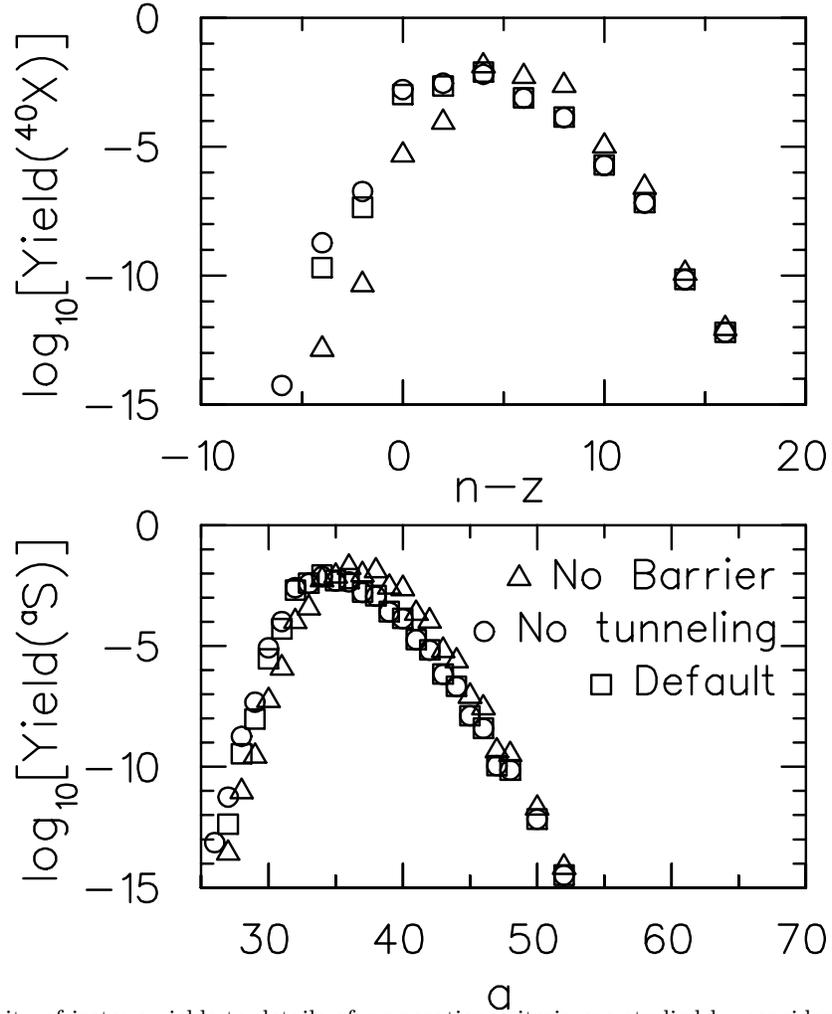}}
\caption{
\label{barrierdependence_fig}
The sensitivity of isotope yields to details of evaporation criteria are
studied by considering the yields of sulphur isotopes from the fragmentation of
an $A=200,~Z=80$ system at a temperature of 4.5 MeV. Results from three cases
are displayed. The evaporative description in the default calculation (squares)
assumes a Coulomb barrier with an $s$-wave tunneling contribution described in
Eq. (\ref{tunneling_eq}). Calculations were repeated with no tunneling
(diamonds) and with the barrier eliminated (triangles), making the emission of
neutrons and protons equal. Yields near the neutron drip line were insensitive
to details of barrier, but yields near the proton drip line were indeed
sensitive to the barrier and to the tunneling.}
\end{figure}

The final aspect of the modeling considered here is barrier penetration, which
plays an important role in sequential decay.  Sequential decay, as seen in
Figs.~\ref{tempdependence_a200z80_a40_fig} and
\ref{tempdependence_a200z80_sulphur_fig}, strongly narrows the isotope
distributions as evaporation pushes the fragments towards the valley of
stability.  The Coulomb barrier also leads to preferential emission of neutrons
relative to protons.  In our calculations a fixed Coulomb barrier was imposed
with an additional consideration of $s$-wave tunneling.  To understand the
importance of tunneling, calculations were performed with and without
tunneling.  Calculations were also performed with the Coulomb barrier
completely absent.  Results considering the fragmentation of the
$A=200,~Z=80$ system at a temperature of 4.5 MeV are displayed in
Fig.~\ref{barrierdependence_fig}.  Ignoring the tunneling increased the yields
of proton-rich fragments but left the yield of neutron-rich fragments
unchanged.  Ignoring the Coulomb barrier altogether significantly reduced the
population of proton-rich fragments and only slightly increased the yield of
neutron-rich fragments.

Interpretation of Fig.~\ref{barrierdependence_fig} suggests that the yield of
neutron-rich fragments is surprisingly insensitive to the imposition of the
Coulomb barrier as the preference of emitting neutrons to protons is more
strongly derived from the greater phase space available when decaying towards
the valley of stability. If one were interested in the production of
proton-rich fragments, there would be a strong motivation to improve the
treatment of the Coulomb barrier.

\section{Isospin amplification and baryonic chemical potentials}
\label{deltamu_sec}

The phenomenon of isospin amplification was presented in
Sec.~\ref{liquiddrop_subsec} in the context of the liquid drop model
and was also discussed in Sec.~\ref{systemdependence_sec}.  It was
seen that relatively small neutron excesses could result in large
increases in the yields of neutron-rich isotopes.  Experimentally, it
has been observed that yields are indeed quite sensitive to the
neutron fraction of the fragmenting system.  By creating a ratio of
yields for different isotopes from the fragmentation of $^{112}$Sn and
dividing by the same ratio for the fragmentation of $^{124}$Sn, one is
able to extract the difference of the chemical potential for the two
fragmenting systems.

To better explain the ratio, one considers the population of a level $i$ of an
isotope of mass $a$ in the grand canonical ensemble,
\begin{equation}
Y(^aX)=\sum_iV(2J_i+1)\left(\frac{amT}{2\pi}\right)^{3/2}\exp\left\{
\frac{-B_i+z\mu_p+(a-z)\mu_n}{T}
\right\}.
\end{equation}
From the results of the previous sections, it is clear that the yield is
significantly affected by sequential decay. However, studies using the MMMC
model \cite{smm_isotoperatios,snisotope_ratios} have revealed a surprising
insensitivity of the following ratio to sequential decay,
\begin{equation}
{\cal R}^{124}_{112}(^aC/^{12}C)=\frac{Y^{124}(^aC)/Y^{124}(^{12}C)}
{Y^{112}(^aC)/Y^{112}(^{12}C)},
\end{equation}
where the 124 and 112 refer to the different isotopes of the overall system.
This ratio should yield $e^{(n-z)\Delta\mu_n/T}$, where $\Delta \mu_n$ is the
difference of the neutron chemical potentials of the $A=124$ and $A=112$
systems. By creating a similar ratio using isotopes of the same neutron number,
one is able to extract $\Delta\mu_p$.

\begin{figure}
\epsfxsize=0.6\textwidth
\centerline{\epsfbox{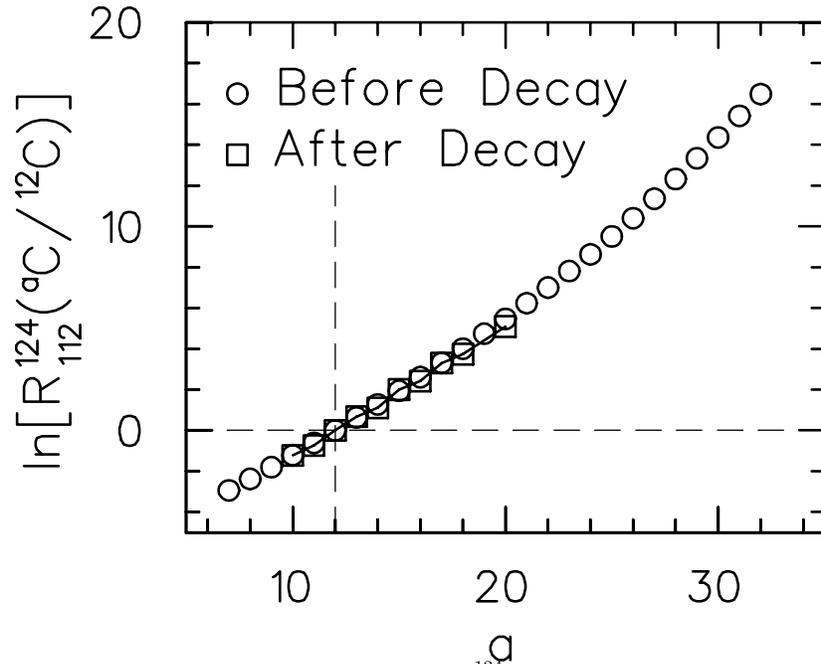}}
\caption{
\label{deltamu_decay_fig}
Ratios of Carbon isotope yields for the fragmentation of $^{124}$Sn are divided
by the same ratios for the fragmentation of $^{112}$Sn to obtain a quantity
which should reflect $\exp\{(n-z)\Delta\mu_n/T\}$, where $\Delta\mu_n$ is the
increase of the neutron chemical potential resulting from using a heavier tin
isotope. The logarithm of that ratio, which is presented here, rises linearly
with $a$ as expected. The pre-decay results (squares) and the post-decay
results (squares) are remarkably similar.}
\end{figure}

To illustrate the insensitivity of the ratio to sequential decay, the logarithm
of the ratio is shown both before and after sequential decay in
Fig.~\ref{deltamu_decay_fig} for the fragmenting $^{112}$Sn and $^{124}$Sn
systems at a temperature of 4.5 MeV. The slight non-linearity of the pre-decay
results derives from the finite size of the system, i.e. a grand canonical
calculation would yield a straight line.  

\begin{figure}
\epsfxsize=0.6\textwidth
\centerline{\epsfbox{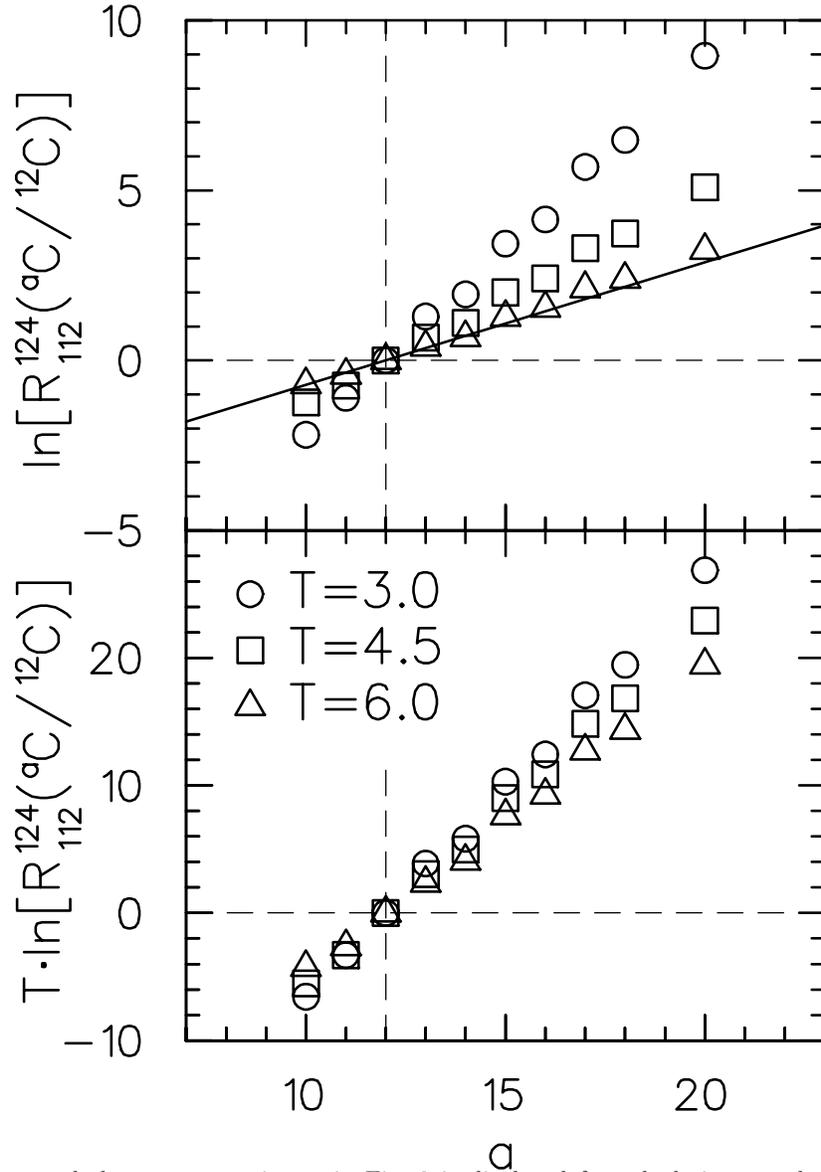}}
\caption{
\label{deltamu_tempdependence_fig}
In the top panel the same quantity as in Fig.~\ref{deltamu_decay_fig} is
displayed for calculations at three temperatures, 3.0, 4.5 and 6.0 MeV.  The
solid line represents the experimentally extracted value of $\Delta\mu_n$ from
reference \protect\cite{snisotope_ratios}. The consistency of the experimental
results with the $T=3$ MeV results may be misleading as the calculations
ignored the fact that some of the extra neutrons associated with $^{124}$Sn
could have evaporated before the onset of multifragmentation. The same results
are displayed in the lower panel, only multiplied by the temperature. This
illustrates that $\Delta\mu_n$ is fairly insensitive to
the temperature.}
\end{figure}

The resulting ratio also clearly depends on temperature and breakup density. To
illustrate the temperature dependence calculations were performed for three
temperatures, 3.0 MeV, 4.5 MeV and 6.0 MeV. The log of the ratio was in this
case multiplied by the temperature so that one could focus on $\Delta \mu$
rather than $\Delta\mu/T$. The slopes weakly depend on the temperature as shown
in the lower panel of Fig.~\ref{deltamu_tempdependence_fig}. The sensitivity to
the breakup density was also investigated and found to be rather weak.

The upper panel of Fig.~\ref{deltamu_tempdependence_fig} displays a line
derived from analysis of several experimentally measured isotope ratios
\cite{snisotope_ratios}, where the ratios were observed to be surprisingly
robust with respect to the choice of element and isotopes.  The comparison with
data is difficult to interpret. At the time in a collision where
multifragmentation occurs, one would expect some of the extra neutrons in the
$^{124}$Sn system to have boiled off. Thus, it is not surprising that the
calculation over-predicts $\Delta\mu_n$ as the calculation assumes the systems
differ by 12 neutrons. The experimental result was also of a different
effective size due to the fact that the mechanism for excitation was
semi-peripheral heavy ion collisions. As the experiment involved $^{124}$Sn +
$^{124}$Sn, $^{112}$Sn + $^{124}$Sn, and $^{112}$Sn + $^{112}$Sn collisions,
the effective system size might have been either more or less than the size of
a single Sn ion depending on the centrality of the collision and experimental
acceptance. Nonetheless, several positive conclusions should be drawn from this
analysis. First, the ratios are indeed robustly independent of sequential
decay, even when one considers isotopes far along the isotope chain from
stability. Secondly, the experimental results would seem to be consistent with
model predictions if one were to assume that approximately a third of the extra
neutrons gained from going from $^{112}$Sn to $^{124}$Sn evaporated prior to
multifragmentation.

\section{Summary and conclusions}
\label{summary_sec}

Statistical multifragmentation has been investigated in regards to the
production of rare nuclides, particularly neutron-rich isotopes. Statistical
calculations have been performed in the canonical ensemble, taking into account
exact conservation of $N$ and $Z$. By comparing to calculations in the grand
canonical ensemble, particle-number constraints were seen to be non-negligible,
especially for the fragmentation of smaller systems. Assuming a simple
liquid-drop form for the ground-state energies and assuming a simplified form
for the distribution and degeneracies of excited states, initial yields were
calculated for every level of every possible isotope. Sequential evaporation
has been modeled by calculating the ways to partition the initial weight of
every particle-unstable level into any state to which it might decay. Eight
modes of decay were considered. Since neither the calculation of the initial
distribution nor the modeling of the sequential decay involved Monte Carlo
procedures, it was possible to calculate yields for extremely rare isotopes.

By considering yields of fragments of a fixed $a$ as a function of $n-z$,
two lessons were learned by considering a simple liquid drop model. First,
the width of the initial distribution was determined largely by the ratio of
the temperature to the symmetry term in the liquid-drop model. Secondly,
isospin amplification, or isospin fractionation, could also be simply
understood in terms of the same two quantities.

Sequential decay dramatically altered the yields, pushing the yields towards
the valley of stability. The effects were strongest for larger fragments and
for higher temperatures. As the initial distributions for fragments of fixed
$a$ as a function of $n-z$ were broadest at high temperature, inclusion of
sequential decay shows that 5 MeV is the best temperature for creating rare
neutron-rich fragments in the $a=40$ region. For heavy fragments with $a=100$,
the optimum temperature was near three MeV which is more in the domain of
fission than the domain of multifragmentation.

In order to understand the sensitivity of the calculations to various aspects
of the modeling, the breakup density, level density, system size and system
neutron fraction were systematically varied. Results were weakly dependent on
the breakup density and moderately sensitive to the level density. Larger
systems lead to somewhat broader yields due to finite-size constraints. Yields
were especially sensitive to small changes in the isospin composition of the
overall system. Thus, choosing projectiles and targets with large neutron
fractions, e.g.\ $N/A$ for Uranium is 0.614 and $N/A$ for $^{124}$Sn is 0.597,
strongly increases the chances of creating isotopes near the neutron drip
line. Finally, the sensitivity of the yields with respect to details of the
evaporation was considered by adjusting the Coulomb barrier that leads to
preferential emission of neutrons as opposed to protons. The yields of
neutron-rich fragments were surprisingly insensitive to the barrier and changed
almost imperceptibly when the Coulomb barrier was removed
altogether. Proton-rich fragments were however quite sensitive to the details,
as the tunneling allowed proton-rich fragments to return to the valley of
stability.

The ability of statistical models to explain isotope production is certainly of
scientific interest in its own right. Additionally, one might also consider
whether multifragmentation could offer a competitive means for creating rare
isotopes. However, separating and focusing particles could be problematic. If
one fragmented heavy nuclei, which have the broadest yields, the Coulomb forces
between the fragment of interest and various parts of the residual system would
spread the emission over a large kinematic region making separation or focusing
of the produced particles difficult. 
This would be especially true for
fragments produced at mid-rapidity. Thus, it is difficult to discern whether
there is a pragmatic side to the isospin degree of freedom in
multifragmentation. To date, experiments that measure isotope yields have
either been designed to focus on projectile rapidities and thus ignore
multifragmentative events, or have measured production of light elements,
$a<20$.  If statistical descriptions are shown to accurately describe isotope
production for a broader range of nuclides, it might warrant serious
consideration of designing an apparatus to capture and identify very rare
isotopes in a very different environment than projectile fragmentation.

\acknowledgements{This work was supported by the National Science Foundation,
grants PHY-96-05207 and PHY-00-0527. Patrick Underhill was supported by the
Research Experience for Undergraduates program at Michigan State University
which is sponsored by the National Science Foundation, grant
PHY-94-24140. Christopher Morling was supported by the Undergraduate
Professorial Assistants Program at Michigan State University. W.B. 
acknowledges support from an Alexander-von-Humboldt Foundation U.S. 
Distinguished Senior Scientist award.}

\newpage

\end{document}